\documentclass[lettersize,journal]{IEEEtran}
\usepackage{amsmath,amsfonts}
\usepackage{algorithmic}
\usepackage{algorithm}
\usepackage{array}
\usepackage[caption=false,font=normalsize,labelfont=sf,textfont=sf]{subfig}
\usepackage{textcomp}
\usepackage{stfloats}
\usepackage{url}
\usepackage{verbatim}
\DeclareUnicodeCharacter{FF5C}{|}
\DeclareUnicodeCharacter{FF5B}{[}
\DeclareUnicodeCharacter{FF5D}{]}
\usepackage{graphicx}
\def\BibTeX{{\rm B\kern-.05em{\sc i\kern-.025em b}\kern-.08em
    T\kern-.1667em\lower.7ex\hbox{E}\kern-.125emX}}
\usepackage{balance}
\begin{document}
\title{CASPER: Contrastive Approach for Smart Ponzi Scheme Detecter with More Negative Samples }

\author{{Weijia Yang, Tian Lan, Leyuan Liu, Wei Chen, Tianqing Zhu, Sheng Wen,  Xiaosong Zhang}
\thanks{Weijia Yang and Tian Lan are with the School of Computer Science and Engineering, University of Electronic Science and Technology of China, Chengdu, Sichuan, China. Leyuan Liu, Wei Chen, and Xiaosong Zhang are with the School of Information and Software Engineering, University of Electronic Science and Technology of China, Chengdu, Sichuan, China. Leyuan Liu is the corresponding author (email:202411081716@std.uestc.edu.cn,
lantian1029@uestc.edu.cn,
leyuanliu@uestc.edu.cn,
chenwei@uestc.edu.cn,
johnsonzxs@uestc.edu.cn).}
\thanks{Tianqing Zhu is with the Faculty of Data Science, City University of Macau, Macao Special Administrative Region, China(email:tqzhu@cityu.edu.mo).}
\thanks{Sheng Wen is with the School of Science, Computing and Emerging Technologies, Swinburne University of Technology, Melbourne, Australia(emial:swen@swin.edu.au).}
}
\newcommand{\orcidauthorA}{https://orcid.org/0009-0006-7561-5095}

\markboth{Journal of \LaTeX\ Class Files,~Vol.~18, No.~9, September~2020}%
{How to Use the IEEEtran \LaTeX \ Templates}

\maketitle

\begin{abstract}

The rapid evolution of digital currency trading, fueled by the integration of blockchain technology, has led to both innovation and the emergence of smart Ponzi schemes. A smart Ponzi scheme is a fraudulent investment operation in smart contract that uses funds from new investors to pay returns to earlier investors. Traditional Ponzi scheme detection methods based on deep learning typically rely on fully supervised models, which require large amounts of labeled data. However, such data is often scarce, hindering effective model training. To address this challenge, we propose a novel contrastive learning framework, CASPER (Contrastive Approach for Smart Ponzi detectER with more negative samples), designed to enhance smart Ponzi scheme detection in blockchain transactions. By leveraging contrastive learning techniques, CASPER can learn more effective representations of smart contract source code using unlabeled datasets, significantly reducing both operational costs and system complexity. We evaluate CASPER on the XBlock dataset, where it outperforms the baseline by 2.3\% in F1 score when trained with 100\% labeled data. More impressively, with only 25\% labeled data, CASPER achieves an F1 score nearly 20\% higher than the baseline under identical experimental conditions. These results highlight CASPER’s potential for effective and cost-efficient detection of smart Ponzi schemes, paving the way for scalable fraud detection solutions in the future.

\end{abstract}


\section{Introduction} \IEEEPARstart{T}he proliferation of smart contract platforms, particularly Ethereum, has significantly reshaped the blockchain ecosystem and the broader digital economy. These platforms enable decentralized applications (DApps) by automating contract execution through blockchain technology, creating a transparent, secure, and immutable environment~\cite{kushwaha2022systematic,zhu1}. However, as blockchain technologies gain widespread adoption, they also attract new forms of malicious activity, especially smart Ponzi schemes, which rely on attracting new investments to pay returns to earlier investors. These schemes have emerged as a significant threat on platforms like Ethereum. A study 
reported that smart Ponzi schemes caused over $\$$600 million in losses nationwide in April 2024. The anonymity provided by blockchain transactions exacerbates the issue, making it difficult to trace fraudulent activities and protect investors~\cite{fan2021spsd,zhu2}.

In the pursuit of safeguarding blockchain users' financial security, researchers have focused on identifying smart contracts that exhibit characteristics of smart Ponzi schemes. Onu et al.~\cite{onu2023detection} employed machine learning techniques to improve the detection efficiency and precision of smart Ponzi schemes. However, their approach relies on the availability of a large dataset of labeled examples for training, which is often challenging to acquire. Ibba et al.~\cite{ibba2021evaluating} introduced a machine learning model that uses text classification metrics to identify smart contracts demonstrating behaviors associated with smart Ponzi schemes. Yet, like Onu et al.'s approach, their method is also dependent on a substantial corpus of labeled data and may require retraining to adapt to new manifestations of smart Ponzi schemes. Liang et al.~\cite{liang2021data} leveraged dynamic graph embedding techniques to autonomously learn representations of accounts from multi-source, multi-modal datasets, achieving promising results. However, the computational resources required and the model’s ability to detect smart Ponzi schemes across diverse scenarios still need further validation. In conclusion, despite the proliferation of machine learning-based detection methods for smart Ponzi schemes, the field faces challenges such as heavy reliance on feature engineering and limited generalizability across different data distributions—issues that practitioners are eager to address.

Additionally, some researchers have turned to deep learning methodologies to tackle this problem.  Wang et al.~\cite{wang2021ponzi} utilized Long Short-Term Memory (LSTM) networks combined with oversampling techniques to address class imbalance, improving the model's ability to recognize minority classes. However, their approach depends on extensive data and computational resources for training, and it is susceptible to overfitting when data is insufficient. Cui et al.~\cite{cui2024ponzi} employed Convolutional Neural Networks (CNNs) and Bidirectional Gated Recurrent Unit (BiGRU) networks to extract spatial and semantic features, capturing semantic information from smart contract opcodes at various levels. They also integrated attention mechanisms to assign different weights to distinct features, enhancing smart Ponzi scheme detection. While deep learning methods can extract features more effectively, they are highly dependent on data annotation, which in practice requires significant human resources and time.

To address these challenges more effectively, we introduce the CASPER (Contrastive Approach for Smart Ponzi scheme detectER with more negative samples) model, a comprehensive smart Ponzi scheme detection system that integrates a self-supervised representation learning module with a semi-supervised classification module. To enhance feature extraction, we propose an innovative contrastive learning framework and introduce a novel multi-vector cosine similarity method. This method incorporates an intermediate vector that maintains consistent angles with all input vectors, effectively reducing sensitivity to extreme values while preserving essential information. This advancement not only improves the model’s robustness but also accelerates convergence during training. Moreover, we validated the model’s performance on new data structures, showcasing its generalizability. The main contributions of this work are as follows:

\begin{itemize}
    \item To better learn feature representations, we crafted a completely new contrastive learning framework.
    \item To address the issue of insufficient labeled data, we designed an smart Ponzi scheme classification model combining self-supervised representation learning and semi-supervised classification, which can perform smart Ponzi scheme identification with fewer labeled data.
    \item We derive and calculate the relationship and method for calculating the cosine similarity between multiple vectors, which can be the basis for implementing a self-supervised representation learning module and improve the performance of representation learning at the same time.
\end{itemize}
\section{preliminary} \label{section: preliminary} 
\subsection{Symbol Table}

\begin{table}[h!]
\caption{Symbol Table}
\centering
\begin{tabular}{cl}
\hline
\textbf{Symbol} & \textbf{Description} \\ \hline
 $ \mathcal{D} $  & Dataset of smart contracts \\ 
 $ \mathcal{D}_L $  & Labeled subset of  $ \mathcal{D} $  \\ 
 $ \mathcal{D}_U $  & Unlabeled subset of  $ \mathcal{D} $  \\ 
 $ \mathbf{x}_i $  &  $ i $ -th smart contract source code \\ 
 $ y_i $  & Label of  $ \mathbf{x}_i $  (1 for Ponzi scheme, 0 otherwise) \\ 
 $ \mathcal{A} $  & Augmentation function \\ 
 $ \mathbf{x}_i^{(k)} $  &  $ k $ -th augmented view of  $ \mathbf{x}_i $  \\ 
 $ \mathbf{z}_i^{(k)} $  & Feature representation of  $ \mathbf{x}_i^{(k)} $  \\ 
 $ \mathcal{L}_{\text{s,m,w}} $  & Contrastive loss function \\ 
 $ \mathcal{L}_{\text{sup}} $  & Supervised loss function \\ 
 $ \mathcal{L}_{\text{pseudo}} $  & Pseudo-label loss function \\ 
 $ \theta $  & Confidence threshold for pseudo-labeling \\ 
 $ \tau $  & Temperature parameter in contrastive learning \\ 
 $ \lambda_1, \lambda_2 $  & Weight coefficients for total loss \\ 
 $ \mathcal{C} $  & Classifier \\ 
 $ \hat{y}_i $  & Predicted label of  $ \mathbf{x}_i $  \\ 
 $ \mathcal{F} $  & Feature extractor (e.g., GraphCodeBERT~\cite{Guo2020}) \\ 
 $ \mathcal{S} $  & Similarity function (e.g., multi-vector cosine similarity) \\ \hline
\end{tabular}

\end{table}

\subsection{Formal Problem Definition}


The framework (Figure~\ref{fig:overall}) is mainly divided into three steps: 1) Contract representation learning: The contract feature representation is learned by a self-supervised method. 2) Semi-supervised classifier: A semi-supervised training strategy is used to train a classifier using a small amount of labeled data. 3) Prediction: The unknown contract is correctly classified as "Ponzi" or "Non-Ponzi" by the pre-trained feature encoder and the classifier working together.
  Given a dataset  $ \mathcal{D} = \{(\mathbf{x}_i, y_i)\}_{i=1}^N $  of smart contracts, where  $ \mathbf{x}_i $  represents the source code of the  $ i $ -th smart contract and  $ y_i \in \{0, 1\} $  indicates whether the contract is a Ponzi scheme (1) or not (0), our goal is to develop an effective detection model that can accurately classify new, unseen smart contracts with limited labeled data.

The dataset is divided into two parts:
\begin{itemize}
    \item Labeled subset  $ \mathcal{D}_L = \{(\mathbf{x}_i, y_i)\}_{i=1}^M $  with  $ M \ll N $ .
    \item Unlabeled subset  $ \mathcal{D}_U = \{\mathbf{x}_i\}_{i=M+1}^N $ .
\end{itemize}

The model inputted the contract source code and obtained its Data Flow Graph (DFG) data, and GraphCodeBert was used to extract features and input them into the classifier to obtain the classification results . The problem can be formulated as follows:

\textbf{Contract Representation Learning}: Learn a feature extractor  $ \mathcal{F} $  that maps each smart contract source code  $ \mathbf{x}_i $  to a feature representation  $ \mathbf{z}_i \in \mathbb{R}^d $ . This is achieved through a contrastive learning framework that leverages multiple augmented views  $ \mathbf{x}_i^{(k)} $  generated by an augmentation function  $ \mathcal{A} $ . The contrastive learning objective is to maximize the similarity between positive pairs (augmented views of the same contract) and minimize the similarity between negative pairs (augmented views of different contracts). The contrastive loss function  $ \mathcal{L}_{s,m,w} $  is defined to achieve this objective.

\textbf{Semi-supervised Classification}: Utilize the labeled data  $ \mathcal{D}_L $  and pseudo-labeled data from  $ \mathcal{D}_U $  to train a classifier  $ \mathcal{C} $ . The total loss function  $ \mathcal{L}_{\text{total}} $  combines the supervised loss  $ \mathcal{L}_{\text{sup}} $  and pseudo-label loss  $ \mathcal{L}_{\text{pseudo}} $ :
    \begin{equation}
        \mathcal{L}_{\text{class}} = \lambda_1 \mathcal{L}_{\text{sup}} + \lambda_2 \mathcal{L}_{\text{pseudo}}
    \end{equation}
\textbf{Prediction}: For a new smart contract  $ \mathbf{x} $ , the model predicts its label  $ \hat{y} $  using the trained classifier  $ \mathcal{C} $:
    \begin{equation}
        \hat{y}=\mathcal{C}(x)=\left\{\begin{matrix}
 ponzi  \\
 non-ponzi
\end{matrix}\right.
    \end{equation}.

By formulating the problem in this manner, we aim to address the challenges of limited labeled data and improve the generalizability of the detection model through a unified contrastive representation learning framework and semi-supervised classification.

\section{Methodolgy} \label{section:method} 
\subsection{Overall Framework}

\begin{figure*}[ht]
    \centering
    \includegraphics[width = 18cm, height = 12cm]{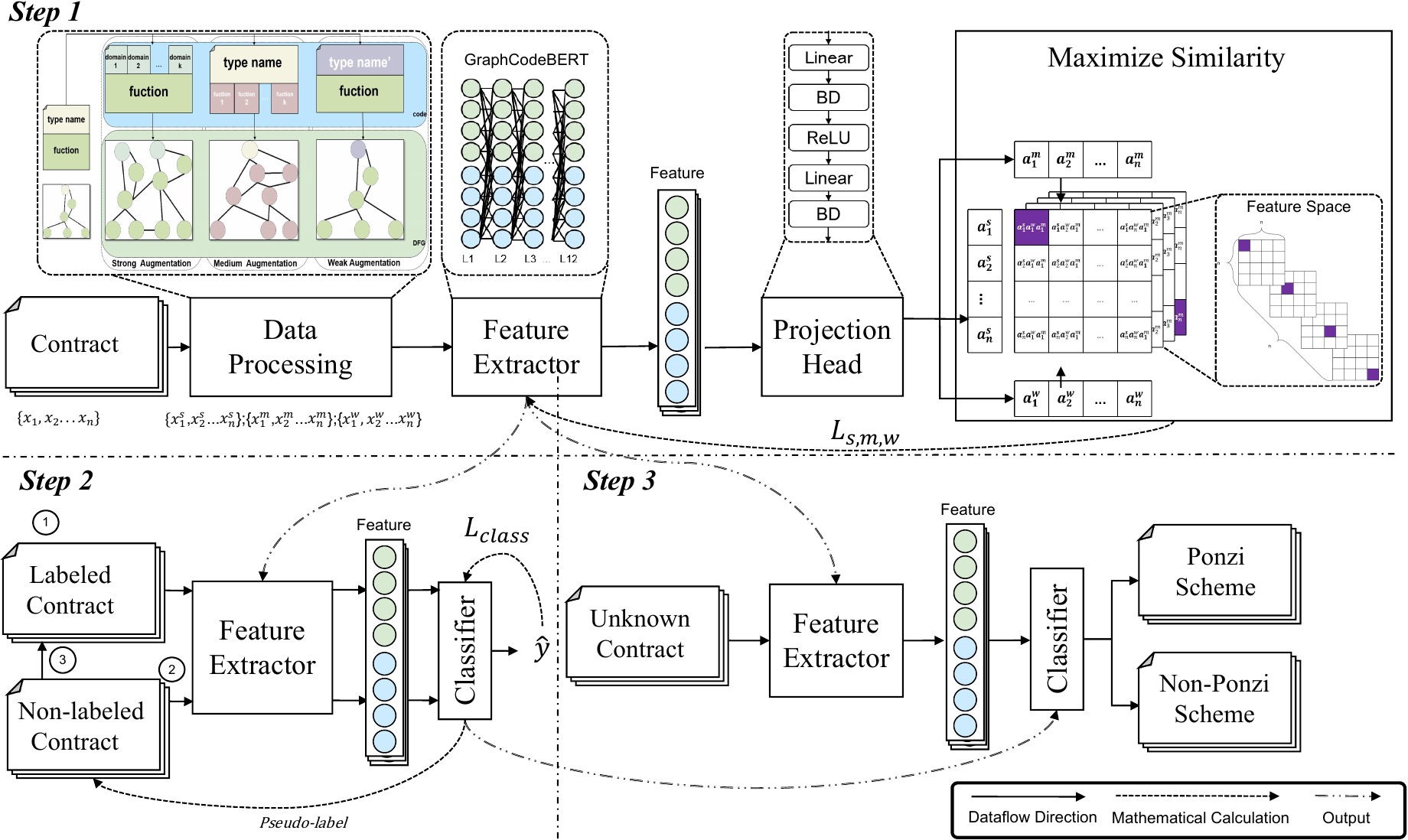} 
    \caption{The overall framework: \textbf{Step 1}: Train the model's feature extractor with a preset self-supervised representation learning framework using unlabeled data. \textbf{Step 2}:The classifier is trained jointly using labeled and unlabeled data by extracting features through a feature extractor. \textbf{Step 3}: Input the unknown smart contract into the trained feature encoder to extract the features and input them into the trained classifier to complete the classification task.}
    \label{fig:overall}
\end{figure*}

The framework of self-supervised representation learning (Figure \ref{fig:overall}) consists of three main components: 1)\textit{Data Preprocessing} : The raw dataset is cleaned and augmented to produce three different views. At the same time, the code features of the three views are extracted and their corresponding DFG views are extracted. 2) \textit{Feature Extractor}: Each view goes through a common feature extraction process to produce a robust feature representation, and the features from different views are mapped into the same feature space by Projection Head. 3) \textit{Maximize Similarity}: The extracted features are arranged into a $n \times n \times n$ matrix, where each axis corresponds to one of the three augmented views. Diagonal elements are positive samples and off-diagonal elements are negative samples. By maximizing the similarity of positive samples and minimizing the similarity of negative samples, the model effectively expands its negative sample pool and enhances the overall representation robustness.

The classifier training takes a simple strategy flow as follows: First, the labeled data is used to obtain features through the trained GraphCodeBERT and a simple linear classifier is used to complete the supervised training. Secondly, the unlabeled data were passed through the same process to obtain features, and the classifier was used to complete the prediction and generate the pseudo-labels of the labeled data. Finally, the pseudo-labels were used to label the unlabeled data and the labeled data were combined to train the classifier together, and the training of the classifier was completed by multiple iterations.

\subsection{Data preprocessing}
Given an input sequence \( x = \{x_1, x_2, \dots, x_n\}\), the first step is to apply data augmentation to obtain three distinct views. 


Strong augmentation involves splitting variables into multiple sub-variables. This method aims to increase the complexity of the code by introducing additional variables that serve similar purposes. The process can be described as follows:

Given a variable declaration  $ x_i $  in the form:
\begin{equation}
    x_i = \text{V}; \text{V}\in \mathbb{H},
    \label{eq:3-1}
\end{equation}
where  V is $\text{variables}$ which is the object used to execute or be passed by the function in the smart contract code and  $\mathbb{H}$is the domain of $\text{variables}$. The strong data augmentation transforms this declaration into:

\begin{equation}
x_i^s = \{\text{V}_1 \in \mathbb{H}_1; \ \text{V}_2 \in \mathbb{H}_2; \ \ldots; \ \text{V}_k \in \mathbb{H}_k\},
\label{eq:3-2}
\end{equation}
where  $ \mathbb{H}_1 \cup \mathbb{H}_2 \cup \ldots \cup \mathbb{H}_k = \mathbb{H}$ and $ k $  is a random integer between 2 and 5. This transformation ensures that the original variable is replaced by  $ k $  sub-variables, each with a unique suffix.

Medium augmentation involves replacing the logic within functions with simpler, predefined logic. This method aims to simplify the code while preserving the overall structure. The process can be described as follows:

Given a function definition  $ x_i $  in the form:

\begin{equation}
x_i = \text{function}\ \text{func}(\text{parameters})\ \text{modifiers}\ \{ \text{body} \},
\label{eq:3-5}
\end{equation}
the weak data augmentation transforms the function body into:

\begin{equation}
x_i^w = \text{function}\ \text{func}(\text{parameters})\ \text{modifiers}\ \{ \text{return}\ \text{value}; \},
\label{eq:3-6}
\end{equation}
where  $\text{value}$  is a simple return value appropriate for the function's return type. For example, if the function returns a boolean,  $\text{value}$  is set to \texttt{true}.

Weak augmentation focuses on renaming variables to introduce variability in the code. This method involves replacing the original variable names with new, randomly generated names. The process can be described as follows:

Given a variable declaration  $ x_i $  in the form:

\begin{equation}
x_i = \text{type}\ \text{name},
\label{eq:3-3}
\end{equation}
the medium data augmentation transforms this declaration into:

\begin{equation}
x_i^m = \text{type}\ \text{name}',
\label{eq:3-4}
\end{equation}
where  $\text{name}'$  is a new variable name generated randomly. The new name is chosen such that it does not conflict with existing variable names in the code. This transformation ensures that all instances of the original variable name are replaced with the new name.

We augmented the source code to obtain data from three views using three data augmentation methods: $x_s = \{x_1^s,x_2^s...x_n^s\}$ , $x_w = \{x_1^w,x_2^w...x_n^w\}$ and $x_m = \{x_1^m,x_2^m...x_n^m\}$

Next, each view’s source code is passed to the Abstract Syntax Tree (AST) and DFG modules to derive corresponding ASTs and DFGs. The AST construction uses the tree-sitter tool~\cite{11} and tree-sitter-Solidity~\cite{100}, originally inspired by tree-sitter-javascript~\cite{13}, to convert Solidity source code into ASTs: $V_s = \{v_1^s,v_2^s...v_n^s\}$ , $V_w = \{v_1^w,v_2^w...v_n^w\}$ and $V_m = \{v_1^m,v_2^m...v_n^m\}$.

Using each AST, a DFG is formed by treating variables as nodes and data dependencies as directed edges. For the strong-augmentation view, edges \(\varepsilon_s = \{v_i^s, v_j^s\}\) indicate dependencies from \(v_i^s\) to \(v_j^s\), with the full edge set denoted by \( E_s = \{\varepsilon_1^s, \varepsilon_2^s, \ldots, \varepsilon_n^s\}\). Thus, the resulting DFG is $G_s = \{V_s, E_s\}$.

Similarly, the weak-augmentation and middle-augmentation views produce \( G_w = \{V_w, E_w\}\) and \( G_m = \{V_m, E_m\}\), respectively.
\subsection{Feature Extractor}

The source code and DFGs for each view are then fed into GraphCodeBERT~\cite{Guo2020} for feature extraction. Taking the strong-augmentation view as an example, the model compiles source code \(S_s\) and DFG \(V_s\) into an input sequence: $\omega_s = \{\texttt{[CLS]}, S_s, \texttt{[SEP]}, V_s\}$.
where \(\texttt{[CLS]}\) is a special classification token, and \(\texttt{[SEP]}\) separates different data types. For each token in \(\omega_s\), the sum of the token embedding and position embedding forms the initial input vector.

For each marker in the sequence $\beta_s$, calculate the sum of its marker embedding and position embedding to obtain the input vector $\omega_{s0}$.
\begin{equation}
    \beta_0^s = TE(\omega_s)+PE(\omega_s),
    \label{eq:3-7}
\end{equation}
where  $TE(\omega_s)$  is the token embedding of the token  $\omega_s$ , and  $PE(\omega_s)$  is the position embedding of  $\omega_s$ . The token embedding is typically a dense vector learned during pre-training, while the position embedding is a fixed or learned vector that encodes the position of the token in the sequence. 

The input vector $\beta_0^s$ is processed through $L$ layers of Transformer, and the output of each layer depends on the output of the previous layer. The output of the nth layer is represented as $\omega_l^s$.
\begin{equation}
    \beta_l^s = transformer^n(\beta_{sl-1}),  l \in [1,L],
    \label{eq:3-8}
\end{equation}
where \(n=12\) follows the GraphCodeBERT configuration. Consequently, the output feature vector from the strong-augmentation view is \(\beta_l^s\). And then We use a MLP with one hidden layer to obtain $\alpha_i^s = g(\beta_l^s) = W^{(2)}\sigma (W^{(1)}\beta_l^s)$  where $\sigma$ is a ReLU nonlinearity.

Analogously, the weak-augmentation and middle-augmentation views yield \(\alpha_l^w\) and \(\alpha_l^m\).

\subsection{Maximize Similarity}
After obtaining the three feature vectors, each is projected into a common feature space for similarity measurement. Various similarity metrics exist (e.g., centroid cosine similarity~\cite{blei2007correlated}, minimal similarity~\cite{karypis1999hierarchical}, weighted average similarity~\cite{ma2008sorec}, variance-based measures), but this work designs a centroid cosine similarity-based approach to favor faster convergence. An intermediate vector \(v\) represents the overall similarity:

\begin{eqnarray}
    sim(\alpha_l^s,\alpha_l^w,\alpha_l^m)=cos<\alpha_l^s,v>,
    \label{eq:3-9}
\end{eqnarray}

Given a batch of $N$input samples, three augmented source codes are generated for each sample, yielding a total of $3N$source code. After importing them into the feature space of the same dimension and pairing them, $N^3$samples are obtained. Among them, the paired sample pairs $(\alpha_l^s,\alpha_l^w,\alpha_l^m)$ from the same sample after enhancement are the positive samples in this batch of samples, the number is $N$, and the remaining paired groups are the negative samples in this batch of samples, the number is $N^3-N$.

For each group, we minimize the similarity loss on the positive samples using the similarity measure we mentioned as a representation of the whole group:
\begin{align}
    L_{s,m,w} = -\frac{1}{N}\sum^N_{i=1}log\frac{exp(sim(\alpha_i^s,\alpha_i^w,\alpha_i^m)/\tau)}
    { \sum_{(m,n,h)\in \mathcal{N}_i}exp(sim(\alpha_m^s,\alpha_n^w,\alpha_h^m)/\tau)},
    \label{eq:3-10}
\end{align}
where, $\tau$ is a temperature parameter, $\mathcal{N}_i$ is the collection of the negative sample group.

The incorporation of a large number of negative samples in our framework significantly enhances the robustness and generalizability of the learned representations. By maximizing the similarity among positive samples and minimizing it among negative samples, our model effectively expands the negative sample pool. This strategy not only improves the model's ability to distinguish between different classes but also reduces the risk of overfitting, leading to better performance on unseen data.

\subsection{Classifier}

In this paper, we use a semi-supervised classifier based on Self-Training with Confidence Thresholding to fully utilize the limited labeled data and a large amount of unlabeled data. The core idea of this method is to iteratively select pseudo-labels with high confidence to expand the training set, thereby enhancing the model's performance.Specifically, we first train an initial classifier using the labeled data. Then, we use this classifier to predict the unlabeled data and select the prediction results with confidence scores higher than a certain threshold as pseudo-labels. These pseudo-labels, together with the labeled data, are used to further train the classifier. This process is repeated until a termination condition is met (e.g., reaching the maximum number of iterations or the pseudo-labels no longer change).The key to this method lies in the selection of the confidence threshold. By setting a reasonable threshold, low-quality pseudo-labels can be effectively filtered out, thus avoiding negative impacts on model training. Moreover, dynamically adjusting the threshold is also an effective means of improving model performance.

We feed the training data used to train the classifier into the representation learning part to obtain the feature representation $F$ of the data and divide it into: $F_{labeled}, F_{unlabeled}$ according to whether it has a label or not. We start by selecting the labeled feature representation F1 to train an initial classifier and compute the loss:
\begin{equation}
   L_{sup}=-\sum^{N}_{i=1}y_ilog(\hat{y_i}),
    \label{eq:3-11}
\end{equation}
where $y_i$ is the true label of the $i -th$ labeled sample, $\hat{y}_i$ is the classifier's predicted output, and N is the number of labeled data.

For the unlabeled data $F_{unlabeled}$ , the classifier's predicted output is $\hat{y}_{unlabeled}$ . We select the prediction results with confidence scores higher than the threshold $\theta$ as pseudo-labels:

\begin{equation}
   S= \{i|max(\hat{y_{unlabeled,i}})\ge \theta\},
    \label{eq:3-12}
\end{equation}
where $S$ is the set of sample indices that meet the confidence threshold.

Thus, the loss of pseudo-label data can be obtained as follows.
\begin{equation}
   L_{pseudo}=-\sum_{j\in S}\hat{y}_{unlabeled,j}log(\hat{y}_{unlabeled,j}),
    \label{eq:3-13}
\end{equation}
where, $\hat{y}_{unlabeled,j}$ is the pseudo-label prediction output of the unlabeled data .

We merge the pseudo-labeled data with the labeled data to form a new training set:
\begin{equation}
   F_{new}=F_{labeled} \cup F_{unlabeled,i},
    \label{eq:3-14}
\end{equation}

\begin{equation}
   y_{new}=y_{labeled} \cup \hat{y}_{unlabeled,i},
    \label{eq:3-15}
\end{equation}
We repeat the above process until a termination condition is met. The loss function for each iteration can be expressed as:
\begin{equation}
   L_{class}=\lambda_1L_{sup}+\lambda_2L_{pseudo},
    \label{eq:3-16}
\end{equation}
where,$\lambda_1$ and $\lambda_2$ are weight coefficients used to adjust the contribution of labeled and pseudo-labeled data in the total loss.

\section{Multi-vector cosine similarity} In Section~\ref{section:method}, several approaches for measuring the similarity among multiple vectors were discussed. Given the specificity of the current task, a method is required that preserves as much information as possible for model learning. Moreover, because the model’s robustness hinges on maximizing the number of negative samples, techniques relying on group-specific maxima or minima may inadvertently reduce the availability of negative samples during training. Through further investigation, methods such as centroid cosine similarity and variance-based measures were identified as potential candidates. However, centroid cosine similarity can be viewed as a disguised average that is sensitive to outliers, and its computational complexity is high when directly applied to our framework. Consequently, the question arises: Can an intermediate vector be found, unaffected by extreme values, while still utilizing the centroid cosine similarity concept for computing an overall cosine similarity among multiple vectors?

\subsection{Method proof}

\begin{figure}[ht]
\centering
\includegraphics[width=6.5cm ,height=3.5cm]{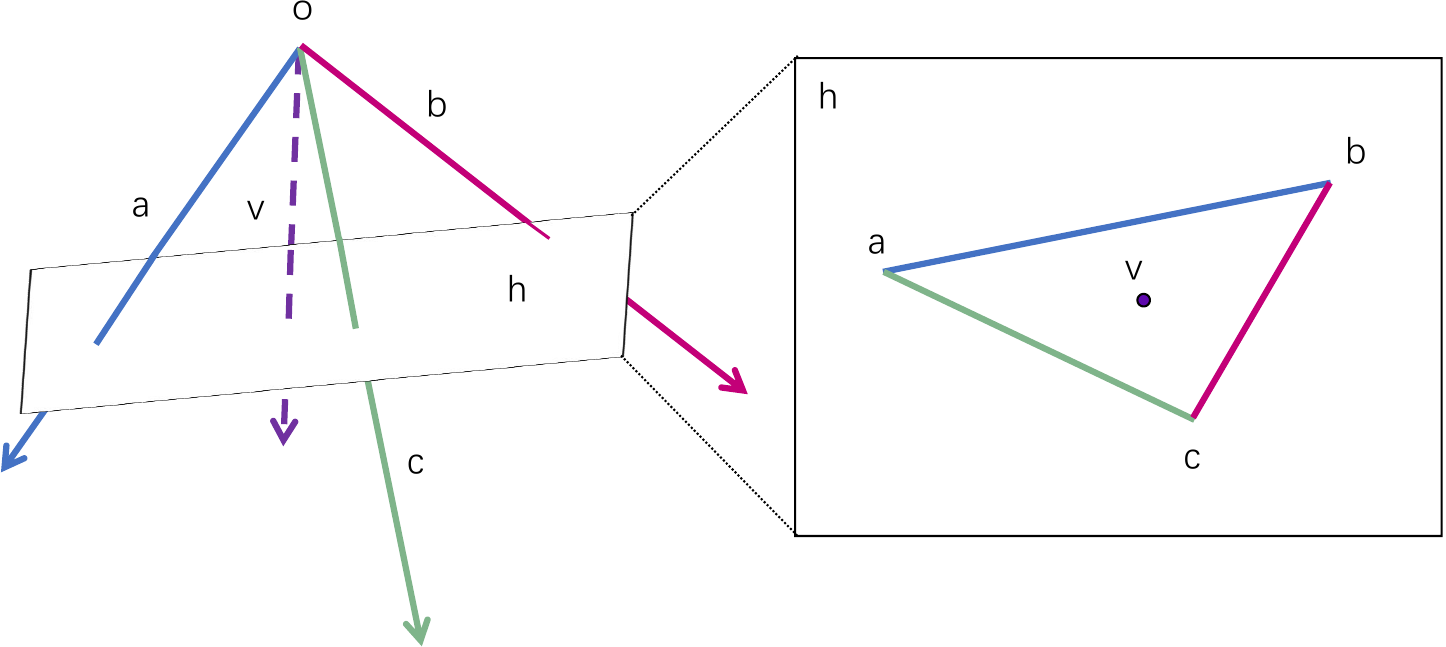} 
\caption{Schematic diagram used to prove the existence of vectors.}
\label{fig:3}
\end{figure}

In three-dimensional space $\mathbb{R}^3$, given three non-coplanar vectors $\overrightarrow{a}$, $\overrightarrow{b}$, and $\overrightarrow{c}$, there exists a vector $\overrightarrow{v}$ such that the angles between $\overrightarrow{v}$ and each of the vectors $\overrightarrow{a}$, $\overrightarrow{b}$, and $\overrightarrow{c}$ are equal (shows in Figure~\ref{fig:3}).

\textbf{Proof by Contradiction}
\begin{enumerate}
    \item \textbf{Assume the Contrary}: Suppose that there does not exist such a vector $\overrightarrow{v}$ that forms equal angles with $\overrightarrow{a}$, $\overrightarrow{b}$, and $\overrightarrow{c}$.
    
    \item \textbf{Construct a Perpendicular Plane}: Let $\overrightarrow{v}$ be an arbitrary non-zero vector. Construct a plane $H$ that is perpendicular to $\overrightarrow{v}$. Since $\overrightarrow{v}$ is arbitrary, such a plane $H$ always exists.
    
    \item \textbf{Translate Vectors to a Common Origin}: Translate the vectors $\overrightarrow{a}$, $\overrightarrow{b}$, and $\overrightarrow{c}$ so that they all originate from a common point $O$.
    
    \item \textbf{Intersections of the Plane with Vectors}: The plane $H$ intersects the vectors $\overrightarrow{a}$, $\overrightarrow{b}$, and $\overrightarrow{c}$ at points $A$, $B$, and $C$, respectively. These points form a triangle $\triangle ABC$ on the plane $H$.
    
    \item \textbf{Circumcenter of the Triangle}: By the circumcircle theorem, any triangle $\triangle ABC$ has a circumcircle with a center $X$ that is equidistant from the vertices $A$, $B$, and $C$, i.e., $|XA| = |XB| = |XC|$.
    
    \item \textbf{Deriving a Contradiction}:
    \begin{itemize}
        \item Since $X$ is the circumcenter of $\triangle ABC$ and lies on the plane $H$, it is equidistant from $A$, $B$, and $C$.
        \item Because $\overrightarrow{v}$ is perpendicular to the plane $H$, the line passing through $X$ and parallel to $\overrightarrow{v}$ is equidistant from $A$, $B$, and $C$.
        \item This implies that $\overrightarrow{v}$ forms equal angles with $\overrightarrow{a}$, $\overrightarrow{b}$, and $\overrightarrow{c}$, as their projections onto $\overrightarrow{v}$ are equal in length.
    \end{itemize}
    
    \item \textbf{Conclusion}: The above derivation contradicts our initial assumption that no such vector $\overrightarrow{v}$ exists. Therefore, the original statement must be true: there exists a vector $\overrightarrow{v}$ such that the angles between $\overrightarrow{v}$ and $\overrightarrow{a}$, $\overrightarrow{b}$, and $\overrightarrow{c}$ are equal.
\end{enumerate}

\subsection{Formula derivation}
Consider three vectors in $\mathbb{R}^3$:  $\overrightarrow{a}(x_a,y_a,z_a)$, $\overrightarrow{b}(x_b,y_b,z_b)$, $\overrightarrow{c}(x_c,y_c,z_c)$, let $\overrightarrow{v}(x_v,y_v,z_v)$ be the intermediate vector of fixed length $l$ such that $\overrightarrow{v}$ has the same angle to $\overrightarrow{a}$, $\overrightarrow{b}$, and $\overrightarrow{c}$. The following system of equations arises from enforcing equal cosine similarities between $\overrightarrow{v}$ and each of $\overrightarrow{a}$, $\overrightarrow{b}$, and $\overrightarrow{c}$:

\begin{equation}
   \begin{cases}
\frac{x_ax_v+y_ay_v+z_az_v}{l\sqrt[]{x_a^2+y_a^2+z_a^2} } =\frac{x_bx_v+y_by_v+z_bz_v}{l\sqrt[]{x_b^2+y_b^2+z_b^2}} \\
\frac{x_bx_v+y_by_v+z_bz_v}{l\sqrt[]{x_b^2+y_b^2+z_b^2} } =\frac{x_cx_v+y_cy_v+z_cz_v}{l\sqrt[]{x_c^2+y_c^2+z_b^2}} \\
x_v^2+y_v^2+z_v^2 = l^2
\end{cases} 
\end{equation}
Given that vectors $\overrightarrow{a}$,$\overrightarrow{b}$,and $\overrightarrow{c}$ are known,their lengths and relationships are also known,that is:
\begin{equation}
    \begin{cases}
\frac{\sqrt[]{x_a^2+y_a^2+z_a^2} }{\sqrt[]{x_b^2+y_b^2+z_b^2}}=\alpha   \\
\frac{\sqrt[]{x_b^2+y_b^2+z_b^2} }{\sqrt[]{x_c^2+y_c^2+z_c^2}}=\beta 
\end{cases}
\end{equation}
From this, Equation 1 can be simplified to:
\begin{equation}
\begin{cases}
(\alpha x_a - x_b)x_v + (\alpha x_a - y_b)y_v + (\alpha x_a - z_b)z_v = 0\\
(\beta x_b - x_c)x_v + (\beta f_b - y_c)y_v + (\beta f_b - z_c)z_v = 0\\
x_v^2 + y_v^2 + z_v^2 = l^2
\end{cases}
\end{equation}
After obtaining the simplified system of equations,we define constants based on the information within them:

\begin{equation}
\begin{cases}
A_1 = \alpha x_a - x_b, \quad A_2 = \alpha y_a - y_b, \quad A_3 = \alpha z_a - z_b\\
B_1 = \beta x_b - x_c, \quad B_2 = \beta y_b - y_c, \quad B_3 = \beta z_b - z_c.
\end{cases}
\end{equation}

Substituting these relations simplifies the system, yielding a set of linear equations:

\begin{align}
\begin{pmatrix}
 A_1  &A_2  &A_3 \\
 B_1  &B_2  &B_3
 \end{pmatrix}
\begin{pmatrix}
 x_v\\
 y_v\\
 z_v
\end{pmatrix} & 
 = 
\begin{pmatrix}
 0\\
 0
\end{pmatrix}
\end{align}

This represents a homogeneous system of linear equations, and we seek to solve for \(x_v, y_v, z_v\). To find the solution, we need to determine the null space of the coefficient matrix.

The solution to this linear system is a vector in the null space of the coefficient matrix. First, we examine the rank of the matrix. If the rank is 2, then the system has only the trivial solution \(x_v = 0, y_v = 0, z_v = 0\). If the rank is less than 2, then there are infinitely many solutions, and the solution space is two-dimensional.

The coefficient matrix is given by:

\begin{equation}
M =
\begin{pmatrix}
A_1 & A_2 & A_3 \\
B_1 & B_2 & B_3
\end{pmatrix}
\end{equation}

Assuming the rank of the matrix is less than 2 (i.e., there is a nontrivial solution), we can solve for the relationship between \(x_k, y_k, z_k\) using Gaussian elimination or other methods. If the rank is 1, the solution can be expressed in a parametric form. Let \(x_k, y_k, z_k\) be linearly dependent, and we can write them as:

\begin{equation}
\begin{pmatrix}
x_v \\
y_v \\
z_v
\end{pmatrix}
=
\lambda
\begin{pmatrix}
C_1 \\
C_2 \\
C_3
\end{pmatrix}
\end{equation}

where \(\lambda\) is a parameter and \(\begin{pmatrix} C_1 \\ C_2 \\ C_3 \end{pmatrix}\) is a basis vector for the null space of the matrix.

From the spherical constraint \(x_v^2 + y_v^2 + z_v^2 = r\), we substitute the above form into this equation:

\begin{equation}
\lambda^2 (C_1^2 + C_2^2 + C_3^2) = l^2
\end{equation}

Thus, \(\lambda\) can be determined as:

\begin{equation}
\lambda = \pm \sqrt{\frac{l^2}{C_1^2 + C_2^2 + C_3^2}}
\end{equation}

The general solution for the system is:

\begin{equation}
\begin{pmatrix}
x_v \\
y_v \\
z_v
\end{pmatrix}
=
\pm \sqrt{\frac{l^2}{C_1^2 + C_2^2 + C_3^2}}
\begin{pmatrix}
C_1 \\
C_2 \\
C_3
\end{pmatrix}
\end{equation}

where \(\begin{pmatrix} C_1 \\ C_2 \\ C_3 \end{pmatrix}\) is a basis vector for the null space of the linear system, and \(\lambda\) is the scaling factor determined by the spherical constraint.
\section{Experiment Results}\label{section:NEXP}To evaluate the excellence and rationality of CASPER, we will design the following thought-provoking questions for experimental setup and provide experimental results:
\begin{itemize}
    \item \textbf{RQ1}: How does CASPER perform as a smart pnzi scheme contract identification model with limited labeled training data?
    \item \textbf{RQ2}: Can the CASPER method achieve good performance on other datasets to demonstrate its generalization ability?
    \item \textbf{RQ3}: What is the performance of self-supervised representation learning in other tasks?
    \item \textbf{RQ4}: Will ablating some modules of the model affect the final prediction result of the model?
    \item  \textbf{RQ5}: Is self-supervised representation learning effective at learning feature representations of the data?
    
\end{itemize}

\subsection{Experiments Setting}
\subsubsection{Datasets}
Two distinct datasets serve different training objectives in CASPER:

\begin{itemize}
    \item \textbf{Self-supervised pre-training dataset:} A corpus of 10,051 smart contracts crawled from Etherscan~\cite{101}, Blockscout~\cite{102}, and Bscscan~\cite{103}. This dataset is used to train the self-supervised representation model.
    \item \textbf{XBlock dataset~\cite{zheng2023}:} A collection of 6{,}498 smart contracts obtained from Etherscan. Each contract was manually classified by reviewing its source code, referencing prior research methodologies. Among these contracts, 318 are identified as smart Ponzi schemes, while the remaining ones are labeled as non-Ponzi. This dataset serves as the training set for the classifier phase, enabling the model to distinguish Ponzi contracts from legitimate ones.
\end{itemize}

For the evaluation of transferability, CASPER was tested on the following datasets besides \textbf{Xblock} dataset:
\begin{itemize}

    \item \textbf{Honeypot smart contract dataset~\cite{torres2019art}}: Honeypot contracts are designed with hidden traps that victimize attackers trying to exploit vulnerabilities. This dataset includes 323 verified honeypot smart contracts in 8 categories, derived from the first major study on honeypot contracts.
    \item \textbf{Phishing smart contract dataset}: Phishing contracts often use proxies and upgrade features to hide and obfuscate, to blind users and steal funds. We collect 225 Phishing
    smart contracts that were flagged as Phish Hack on Etherscan, combined with 2,122 normal contracts, forming a comprehensive phishing dataset.
\end{itemize}

To evaluate the generalization ability of CASPER, several datasets were used, including:
\begin{itemize}
    \item \textbf{EPSD} \cite{galletta2024explainable}, a labeled dataset consisting of 4422 Ethereum smart contracts, where 3749 (84.78\%) are non-Ponzi schemes and 673 (15.22\%) are smart Ponzi schemes.
    \item \textbf{EBD} \cite{16}, which contains 200 smart Ponzi scheme contracts and 3580 non-Ponzi contracts.
\end{itemize}

\subsubsection{Baselines}
We conduct experiments on several baselines:
\begin{itemize}
    \item \textbf{Ridge-NC~\cite{zheng2023}}: Ridge-NC is a regularized linear regression classifier using N-gram count features to transform text data, capturing syntactic patterns for classification.
    \item \textbf{SVM-NC~\cite{zheng2023}}: SVM-NC uses a Support Vector Machine classifier with N-gram count features, leveraging SVM’s effectiveness in high-dimensional spaces for text classification.
    \item \textbf{XGBoost-TF-IDF~\cite{zheng2023}}: XGBoost-TF-IDF combines XGBoost, a gradient boosting framework, with TF-IDF features to classify text, where TF-IDF quantifies word importance in documents.
    \item \textbf{MulCas~\cite{zheng2023}}: MulCas is a multi-view cascade model that combines complementary information from multiple views to enhance classification performance.
    \item \textbf{SadPonzi~\cite{7}}: SadPonzi is a semantic-aware detection system for identifying smart Ponzi schemes in Ethereum contracts, using heuristic-guided symbolic execution to extract semantic information from contract paths.
    \item  \textbf{SourceP~\cite{20}}: SourceP detects smart Ponzi schemes in Ethereum contracts by analyzing source code and data flow, converting the code into a graph and applying a pre-trained model for identification.
\end{itemize}

\subsubsection{Evaluation Metrics}
We retained the source code of the dataset and included the corresponding index idx and label. If the label value is 1, it indicates that it is a smart Ponzi scheme; if it is 0, it indicates that it is not. Evaluation metrics: In the experiments of this paper, we will evaluate the model's performance using common F1 scores, Precision, and Recall. The calculation method is as follows: we will divide the results of every model prediction into four categories: true positives (TP), false positives (FP), false negatives (FN), and true negatives (TN), and the corresponding confusion matrix is shown in Table\ref{tab:1}.
\begin{table}[h]
\centering
\caption{Predicting the Confusion Matrix for Sample Classes}
\begin{tabular}{ccc}
\hline
        & Positive & Negative \\ \hline
Positive & TP       & FP       \\
Negative & FN       & TN       \\ \hline
\label{tab:1}
\end{tabular}
\end{table}
Based on this, we can derive the calculation formulas for precision (Pre) , recall (Rec) and F1 score:
\begin{equation}
    Pre = \frac{TP}{TP+FP},
    \label{eq:11}
\end{equation}
\begin{equation}
    Rec = \frac{TP}{TP+FN},
    \label{eq:12}
\end{equation}
\begin{equation}
    F1 = 2\times \frac{Pre\times Rec}{Pre + Rec},
    \label{eq:13}
\end{equation}

At the same time, in some experiments, we also used accuracy (Acc) as one of the evaluation metrics, which is calculated as follows:
\begin{equation}
    Acc = \frac{TP + TN}{TP+FP+FN+TN},
    \label{eq:14}
\end{equation}

The subsequent experiments were presented in a way that makes it easy to read, with the results expressed as the original values multiplied by 100.

\subsubsection{Prameters}
The experimental parameters involved in CASPER are as follows: The gradient accumulation steps are set to 1, the learning rate is 1e-5, the weight decay coefficient is 0.0, the epsilon value for the Adam optimizer is 1e-8, the maximum gradient norm is 1.0, the maximum number of steps is -1, warmup steps are 0, the control parameter $\lambda_1$ is 1.0 $
\lambda_2$ is 0.85, $\tau$ is 2 , $k$ is 4. When training the classifier, a dataset split of 60\% for training, 20\% for testing, and 20\% for validation was used.

\subsection{Experiments Result}
\subsubsection{performance of CASPER(RQ1)}
We trained CASPER on the XBlock dataset with label settings of 25\%,50\%,75\%,and 100\% for learning.To the best of our knowledge,we do not have any other methods as a baseline for comparison,so we used the SourceP method,which we mainly referred to,to train it with the same training settings and obtain the validation results as the experimental object for our baseline comparison.To better demonstrate the feature learning effects of our self-supervised representation learning module,we compared CASPER with the current state-of-the-art methods according to the comparative method proposed by Lu et al.~\cite{Lu2024} using the same division of the XBlock dataset.Specifically,we sorted all contracts by the block height at which smart contracts were created and trained the model using the first 250 smart Ponzi contracts and the middle non-Ponzi smart contracts.The test set contains contracts 251-341 and the remaining non-Ponzi smart contracts.Therefore,the training set contains a total of 5,990 smart contracts,while the test set contains 508 smart contracts.Compared with a random division,this division can provide better model performance,i.e.,when the model only has early smart Ponzi contract data,it can detect emerging smart Ponzi contracts.The comparison models include: RidgeNC, SVM-NC,XGBoost-TF-IDF, MulCas and SadPonzi.The first three methods use features extracted from smart contract machine code,while MulCas adds developer features on top of that,and SadPonzi detects smart Ponzi schemes based on the bytecode of smart contracts. CASPER was mainly compared with SourceP,which only needs to use the source code of smart contracts as features to identify smart contract code and reverse identify the source code of smart Ponzi schemes.The final results of the supervised model comparison are presented in Table\ref{tab:Baseline}:
\begin{table}[h!]
\centering
\caption{ Baseline comparison experiments with various models on the XBlock dataset.}
\begin{tabular}{cccc}
\hline
method                 &Precision     &Recall       & F1   \\ \hline
Ridge-NC               &73.1          &45.4         &56.0 \\
SVM-NC                 &73.2          &50.0         &59.4 \\
XGBoost-TF-IDF         &51.6          &39.8         &44.9   \\
SadPonzi               &52.0          &48.8         &50.3   \\
MulCas                 &82.9          &69.9         &75.8 \\
SourceP(25\%)        &84.1          &75.0         &79.3\\
SourceP(50\%)        &88.9          &80.0         &84.2\\
SourceP(75\%)        &89.3          &86.2         &87.7\\
SourceP(100\%)       &89.1          &91.6         &90.3  \\
\textbf{CASPER(25\%) }  &90.4          &94.5         &92.4 \\
\textbf{CASPER(50\%) }  &91.9          &95.4         &93.6 \\ 
\textbf{CASPER(75\%) }  &\textbf{96.2} &92.6         &94.3 \\ 
\textbf{CASPER(100\%)}  &94.1          &\textbf{96.2}&\textbf{95.2} \\ \hline
\end{tabular}
\label{tab:Baseline}
\end{table}

From the above table, we can see that when we train CASPER with the same label setting as the SourceP method for learning, our method achieves higher F1 scores and recall rates than the other baseline methods in all training settings. Notably, when we use 25\% of the data labels for training, our F1 score is already better than the results obtained by SourceP using 100\% data labels. These data indicate that our method, despite using fewer labeled data, exhibits relatively excellent performance. 


\begin{figure}[ht]
        \centering
	\includegraphics[width=\columnwidth]{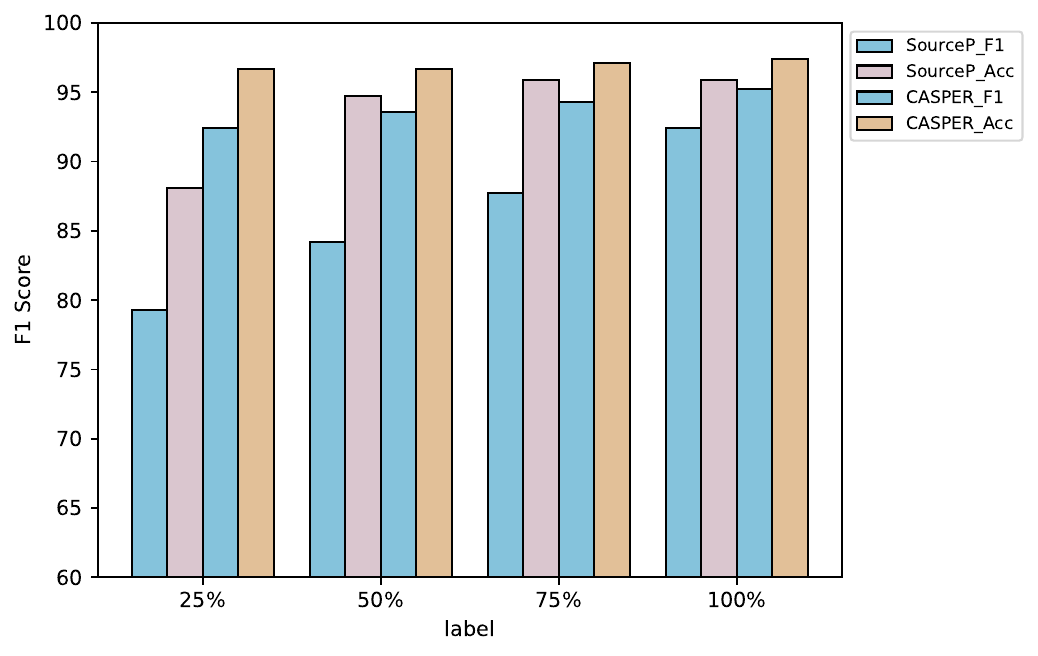} 
	\caption{A bar chart depicting the F1 score and Acc results between CASPER and the SourceP model across different training configurations.}
	\label{fig:baseline}
\end{figure}

From Figure \ref{fig:baseline}, it is evident that CASPER outperforms the SourceP method in both settings. However, an analysis of the trend chart from our experimental results reveals that the performance of the SourceP method is unstable across different experimental configurations. For instance, its F1 score is higher with a 25\% label setting compared to a 50\% label setting. In contrast, CASPER maintains consistent performance across various experimental setups. Furthermore, as a supervised model, the SourceP method exhibits a notable increase in F1 score when trained with 75\% labeled data and under fully supervised training conditions. On the other hand, although CASPER showed better performance than SourceP when transitioning from training with partial labels to training with full labels, the F1 score did not improve significantly. These findings not only shed light on the comparative performance of models in few labels training scenarios but also provide insights into their behavior across varying degrees of supervision.

\subsubsection{Generalization Study (RQ2)}

We conducted experiments on two additional datasets: EPSD and EBD. The validated results are shown in Table \ref{tab:generalization}.
\begin{table}[ht]
    \centering
    \caption{Experimental evaluation of the generalization performance of CASPER across various datasets with 25\% labels.}
    \begin{tabular}{ccccc}
    \hline
    Dataset & Model   & Precision & Recall & F1 Score \\ \hline
    & SVM-NC  &60.9&43.4& 50.6\\
    EPSD& SourceP & \textbf{85.6}& 77.8& 81.5\\
    & CASPER &80.2&\textbf{93.6}  & \textbf{87.7}      \\ \cline{1-5}
    & SVM-NC  & 77.2 & 82.4   & 79.7     \\
    EBD& SourceP & 95.2      & 98.8   & 96.9   \\
    & CASPER&\textbf{99.3}&\textbf{99.7}&\textbf{99.5}        \\ \cline{1-5} 
    \end{tabular}
    \label{tab:generalization}
\end{table}

The performance of CASPER on various datasets indicates that we still achieved good results on the EPSD and EBD datasets in few labels training, demonstrating the strong generalization capability of CASPER.

In addition to validating the model's generalizability through different datasets,we also want to verify our model's performance when faced with unknown data types. We refer to the classification method by Feng, et al. ~\cite{feng2024idponzi} to categorize the smart Ponzi schemes in the dataset into four types: Tree Scheme (\textit{\textbf{TR}}), Chain Scheme (\textit{\textbf{CH}}), Waterfall Scheme 
(\textit{\textbf{WA}}), and Handover Scheme (\textit{\textbf{HA}}). The details of these schemes are as follows:

\begin{itemize}
\item \textit{\textbf{TR}}: utilizes a hierarchical structure where investors join by providing an inviter's address, with earnings distributed hierarchically. The model depends on continuous recruitment of new participants, making it highly prone to collapse.
\item \textit{\textbf{CH}}: follows a linear structure where early investors are rewarded first, while later participants face higher risks of loss. The design is inherently unsustainable due to the sequential payout system.
\item \textit{\textbf{WA}}: distributes returns in a layered manner, favoring early entrants, while later participants receive diminishing returns. Over time, the scheme's financial stability weakens, increasing the risk of failure.
\item \textit{\textbf{HA}}: operates with a simple cyclic structure, where the required investment increases exponentially. Despite its apparent transparency, this rapid escalation in investment demands leads to collapse when the system can no longer sustain payouts.
\end{itemize}

We use a subset of these data types to train the model. The remaining data types are used to evaluate the performance of the model on other types. Meanwhile, we use this experimental setup to train both SVM-NC and SourceP as our comparison models. The specific experimental results are given in Table \ref{tab:generalization-ponzi-type}.

Due to the uneven distribution of data across the four types in the experiment, we apply data augmentation to control the ratio of each data type to 1:1:1:1, and train the model using the augmented dataset. The model's training set consists of the training data and part of the Non-Ponzi data, while the test set consists of the test data and the remaining Non-Ponzi data.

\begin{table}[ht]
\centering
\caption{The experiment for verifying the generalization performance of the model with 100\% labels in the detection of smart Ponzi schemes.}
\begin{tabular}{cccc}
\hline
Training Set&Test Set&Model &F1 \\ \hline
 & &SVM-NC & 48.2\\ 
\textit{\textbf{HA}}  &\textit{\textbf{TR}} $\cup$ \textit{\textbf{CH}} $\cup$ \textit{\textbf{WA}}& SourceP  &61.7\\
 & &CASPER & \textbf{85.9} \\ \cline{1-4}
 & &SVM-NC& 36.5\\
\textit{\textbf{TR}}   &\textit{\textbf{CH}} $\cup$ \textit{\textbf{WA}} $\cup$ \textit{\textbf{HA}}&SourceP   &57.1 \\
 & &CASPER & \textbf{82.5}\\  \cline{1-4}
 & &SVM-NC &28.3  \\
\textit{\textbf{WA}}   &\textit{\textbf{TR}} $\cup$ \textit{\textbf{CH}} $\cup$ \textit{\textbf{HA}}&SourceP   &62.0 \\
 & &CASPER &\textbf{80.5}  \\\cline{1-4}
 & &SVM-NC &44.3  \\
\textit{\textbf{CH}}   &\textit{\textbf{TR}} $\cup$ \textit{\textbf{WA}} $\cup$ \textit{\textbf{HA}}&SourceP  &74.2 \\
 & &CASPER & \textbf{88.7} \\  \cline{1-4}
\hline
\end{tabular}
\label{tab:generalization-ponzi-type}
\end{table}

As shown in Table \ref{tab:generalization-ponzi-type}, our method achieves better performance compared with other baseline models. When the model is trained on a single scheme type and tested on other scheme types, performance remains strong, with a minimum F1 score of 80.5\%. It is worth noting that when training on \textit{\textbf{WA}} and testing on combinations of other scheme types (\textit{\textbf{TR}}, \textit{\textbf{CH}}, \textit{\textbf{HA}}), We achieved an F1 score of 88.7\%. These results show that CASPER can maintain high accuracy even when faced with untrained data of the same task type.

\subsubsection{Transferability Study (RQ3)}
To validate whether our self-supervised representation learning module has adequately learned the fundamental representations of the contract source code, we verified the capability of our work for fraud contract identification on various datasets of fraudulent contracts. At the same time, the SVM-NC model and SourceP model are selected as baselines, while fixing the parameters of their feature extraction module, retraining their classifier parameters under the same dataset, and performing transfer experiments.

\begin{table}[H]
\caption{Transfer experiments in various datasets with 100\% labels.}
\begin{tabular}{ccccc}
\hline
Dataset                                                               & Model   & Precision & Recall & F1 Score \\ \hline

 & SVM-NC  & 95.2      & 43.5   & 59.7     \\
Honeypot  Dataset                                             & SourceP & 83.2      & \textbf{95.2}   & 88.8   \\
                                               & CASPER    &\textbf{97.7} &89.0&\textbf{93.1}        \\ \cline{1-5} 
 & SVM-NC  & 81.8      & 42.4   & 55.8     \\
Phishing   Dataset                                            & SourceP & 88.9      & 85.1   & 87.0    \\
                                               & CASPER     &\textbf{89.8} &\textbf{96.1} &\textbf{92.8}        \\ \hline
\end{tabular}
\label{tab:Transferability Study}
\end{table}

As can be seen in Table \ref{tab:Transferability Study}, we achieve good performance on each fraud dataset compared to our baseline comparison model. These data are enough to show that the representation learning module we designed can better extract the semantic representation into the source code and improve the classification performance on different tasks.


\subsubsection{Ablation experiment(RQ4)}In order to verify the importance of each module of our work setup, we set up a series of ablation experiments to verify: a. the adaptability of the new similarity method to our work; b. impact of DFG, Source Code, and semi-supervised classifier on the final performance of the model.

\textit{a. Compare with similarity method}:  
This experiment we want to fully demonstrate the importance of the similarity calculation method we demonstrated in our work. We verify the performance of this method from two dimensions: one is the performance performance after model training, and the other is the training time required for model training. We use centroid cosine similarity(CCS) and weighted average similarity(WAS) as the baseline methods of our method, train the representation learning part with the same training parameters and perform classification validation with the same classifier.

\begin{figure}[h]
        \centering
	\includegraphics[width=\columnwidth]{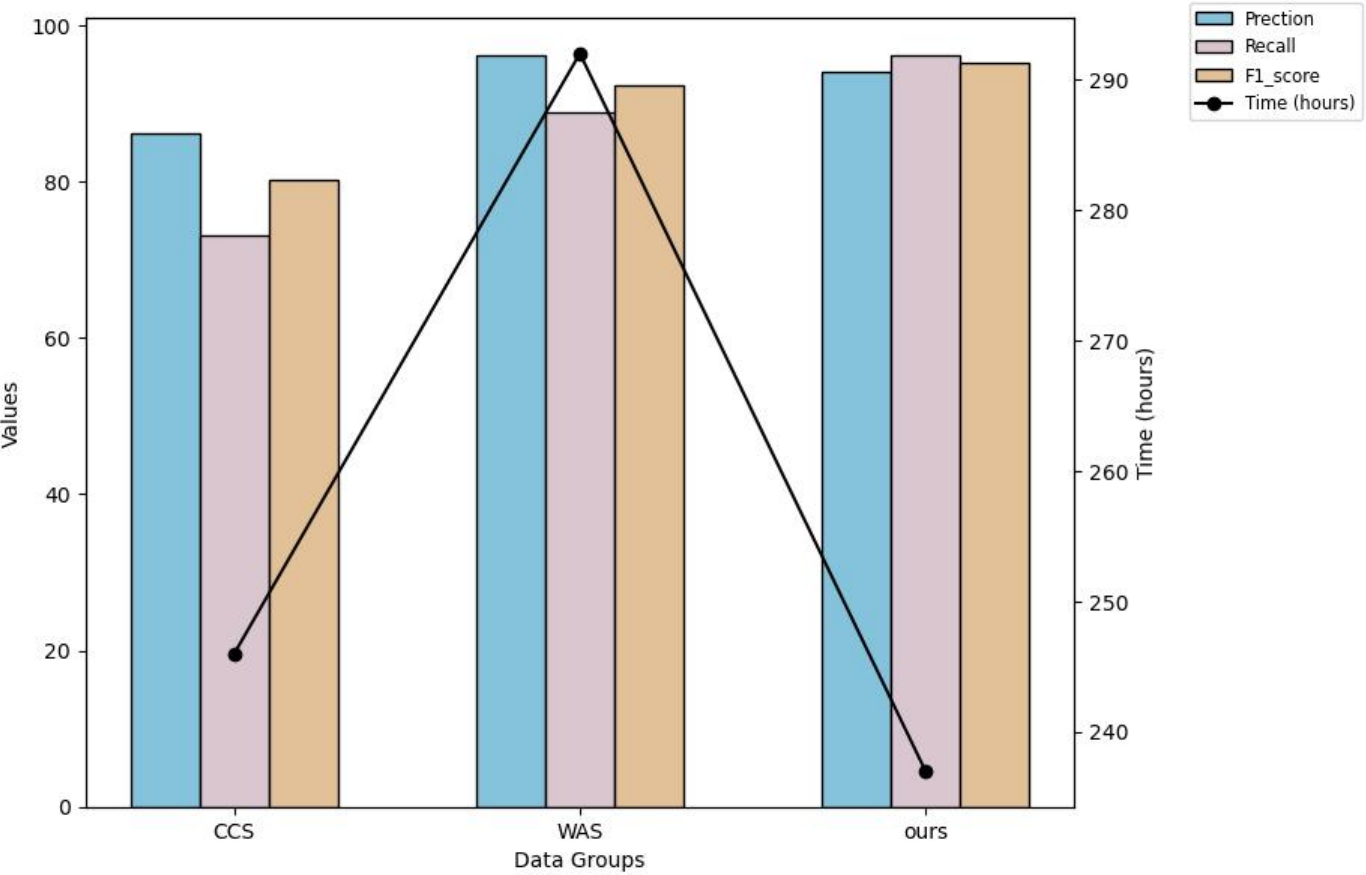} 
	\caption{Accuracy and training time for various similarity calculation methods (The unit of Values is \%, and the unit of Time is hours).}
	\label{fig:similarity}
\end{figure}

From figure \ref{fig:similarity}, we can find that compared with CCS and WAS methods, our method achieves good performance in Recall and F1 score, and uses less training time. Although the Precision of our method is slightly lower than that of the WAS method, our method saves more than 50 hours in training time and has higher overall performance performance. It is enough to show that the similarity calculation method we verified has better performance on our work.

\textit{b. Ablation experiments on DFG, Source Code and classifiers}:
In order to better verify the influence of DFG, Source Code and classifier on the final performance of the model, we designed an ablation experiment to verify the final performance of the model when only 25\% of data labels are used for training under the condition of controlling each input and selecting different classifiers.

\begin{figure}[h]
        \centering
	\includegraphics[width=\columnwidth]{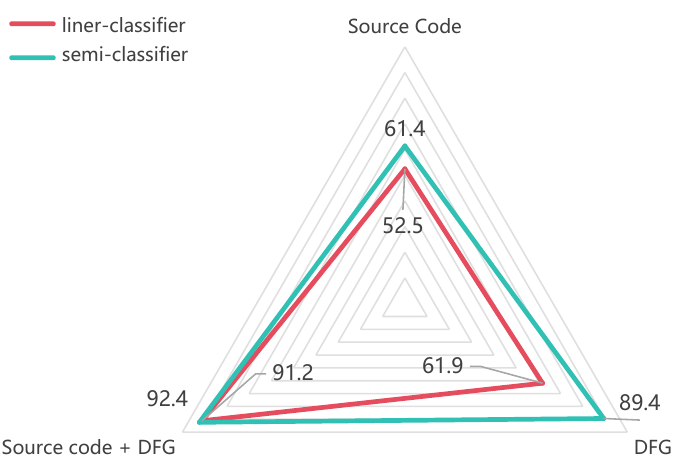} 
	\caption{Radar graphs of ablation experiments for each module of the model.}
	\label{fig:class}
\end{figure}

From the figure \ref{fig:class}, we can see that when using 25\% of the training labels, the performance of the linear classifier only is lower than the performance of the semi-supervised classifier in every scenario. At the same time, we can find that when only DFG is used as input, the semi-supervised classifier improves the final performance of the model greatly, and the model has better performance when only DFG is used as input. It can be inferred that in the task of smart Ponzi scheme classification, the context transfer between functions is more important than the information of the source code itself. The better performance obtained by using DFG and source code as the final input shows that the source code also contains irreplaceable information that can be used in smart Ponzi scheme detection, and this information can be effectively captured and utilized by introducing DFG to finally complete the detection task of smart Ponzi scheme.

\subsubsection{Validation of Representation Learning Performance(RQ5)}
In order to better verify the performance of the representation learning module of our model, we choose different classifiers: SVM, eXtreme Gradient Boosting\cite{chen2016xgboost} (XGBoost) and Mulitilayer perceptron\cite{taud2017multilayer} (MLP) are selected as our downstream classifiers, and the SourceP model and SadPonzi are selected as our comparison object to verify the performance of representation learning of our model under the condition of training the model with 100\% labels.

\begin{table}[ht]
    \centering
    \caption{Comparative Experiments on the Effects of Representation Learning.}
    \begin{tabular}{ccccc}
    \hline
    Dataset & Model   & Precision & Recall & F1 Score \\ \hline
    &SadPonzi & 51.5& 71.2 & 59.8\\
    SVM& SourceP &90.8&95.2&92.9 \\
    & CASPER & \textbf{91.8}&\textbf{96.2}&\textbf{94.0}     \\ \cline{1-5}
    &SadPonzi &78.6& 55.9& 65.3\\
    XGBoost& SourceP &92.0&\textbf{93.1}&92.6    \\
    & CASPER&  \textbf{95.2}&90.8&\textbf{93.0}      \\ \cline{1-5} 
    &SadPonzi &71.7& 55.1&62.3\\
    MLP& SourceP & \textbf{93.6}&92.6&93.1   \\
    & CASPER& 93.0&\textbf{93.5}&\textbf{93.2}       \\ \cline{1-5} 
    \end{tabular}
    \label{tab:representation}
\end{table}

According to Table \ref{tab:representation}, we can find that the overall performance of CASPER is better than that of SourceP model when multiple classifiers are selected. The Recall value of the model using XGBoost is slightly lower than that of SourceP, which may be due to the fact that CASPER learns more rich information when learning data representation, and this information is used for noise representation in the perspective of XGBoost, which slightly influences the recall rate. At the same time, the Precision of the model in the case of using MLP as the classifier is slightly lower than SourceP, which may be due to the fact that a better classification threshold is not obtained during the training process. However, in general, since CASPER has achieved better F1 in the case of using various classifiers, it shows that CASPER has better robustness and the overall performance is better and more stable, and it is suitable for more classifier scenarios.

At the same time, in order to verify the improvement of the representation learning performance by introducing more negative samples, we choose the classical contrastive learning frameworks SimCLR\cite{Chen2020} ( $2n-1$ negative samples) and CLIP\cite{radford2021learning} ( $n^2-n$ negative samples) as the baseline of our experiment. On the Xblock dataset, The representation learning part is trained with the same experimental setup and validated with a fully supervised linear classifier.
\begin{table}[h]
    \centering
    \caption{Table of the performance of the self-supervised representation learning module trained on 10051 samples and a fully supervised classifier under different contrastive learning frameworks.}
    \begin{tabular}{cccc}
    \hline
        Model           &Precision &Recall &F1 score  \\\hline
        Simclr          & 71.4     &89.6   &84.3\\
        CLIP            & 88.5     &93.2   &91.8\\
        CASPER &\textbf{96.6}     &\textbf{93.8}   &\textbf{95.3}\\\hline
    \end{tabular}
    \label{tab:contrastive}
\end{table}

From Table ~\ref{tab:contrastive}, we can find that in the case of self-supervised representation learning, with the increase of the number of negative samples, the model can better learn the representation of data and have better performance in the subsequent classification process.
\section{Related work} \begin{table*}[h!]
\centering
\caption{Summary of Smart Ponzi Scheme Detection Methods by Classification and Method Type}
\begin{tabular}{cp{3cm}p{3cm}p{3cm}p{3cm}}
\hline
\textbf{Method Type \ Classification} & \textbf{User Behavior} & \textbf{Data Flow} & \textbf{Bytecode and Opcode} & \textbf{Source Code} \\
\hline
\textbf{Traditional} & 
Mainly based on transaction patterns ~\cite{bartoletti2020dissecting} and traditional behavior analysis of DApps ~\cite{Cai2018}. Effective for detecting known Ponzi schemes but struggles with emerging patterns. & 
Depends on static analysis of state transitions and data flows, but struggles with complex smart contract logic ~\cite{chen2018detecting}. & 
Traditional techniques such as static bytecode analysis can accurately detect fraudulent behaviors, but are challenged by the complexity and diversity of modern smart contracts ~\cite{Mason2018}. & 
Traditional methods rely on direct analysis of code logic, but struggle with code variants and complex logic ~\cite{Mohanta2018}. \\
\hline
\textbf{Machine Learning} & 
Uses supervised learning models to analyze transaction patterns and user behavior, improving detection accuracy but relying on large labeled datasets ~\cite{Chen2021}. & 
Extracts account and code features and uses algorithms like XGBoost to build detection models, relying on high-quality feature engineering and labeled data ~\cite{chen2018detecting}~\cite{Wang2021}. & 
Employs models like random forests to improve detection performance, but relies on the quality of extracted features ~\cite{Chen2020}. & 
Uses machine learning techniques to analyze source code logic, but struggles with code variants and complex logic ~\cite{Mohanta2018}. \\
\hline
\textbf{Deep Learning} & 
Emphasizes dynamic analysis and real-time user behavior monitoring, but relies on large-scale transaction data and struggles with emerging schemes ~\cite{Chen2021}. & 
Uses deep learning models such as TextCNN and Transformer to extract structured information, but with high computational overhead and challenges in real-time deployment ~\cite{chen2021TextCNN}. & 
Employs semantic analysis of bytecode to improve detection performance, but struggles with novel fraud patterns ~\cite{7}. & 
Leverages pre-trained code representations (e.g., GraphCodeBERT) and data flow analysis to enhance feature extraction, capturing fraudulent behaviors more comprehensively ~\cite{Guo2020}. \\
\hline
\end{tabular}
\label{tab:related}
\end{table*}

\begin{figure}[h]
    \includegraphics[width=\columnwidth]{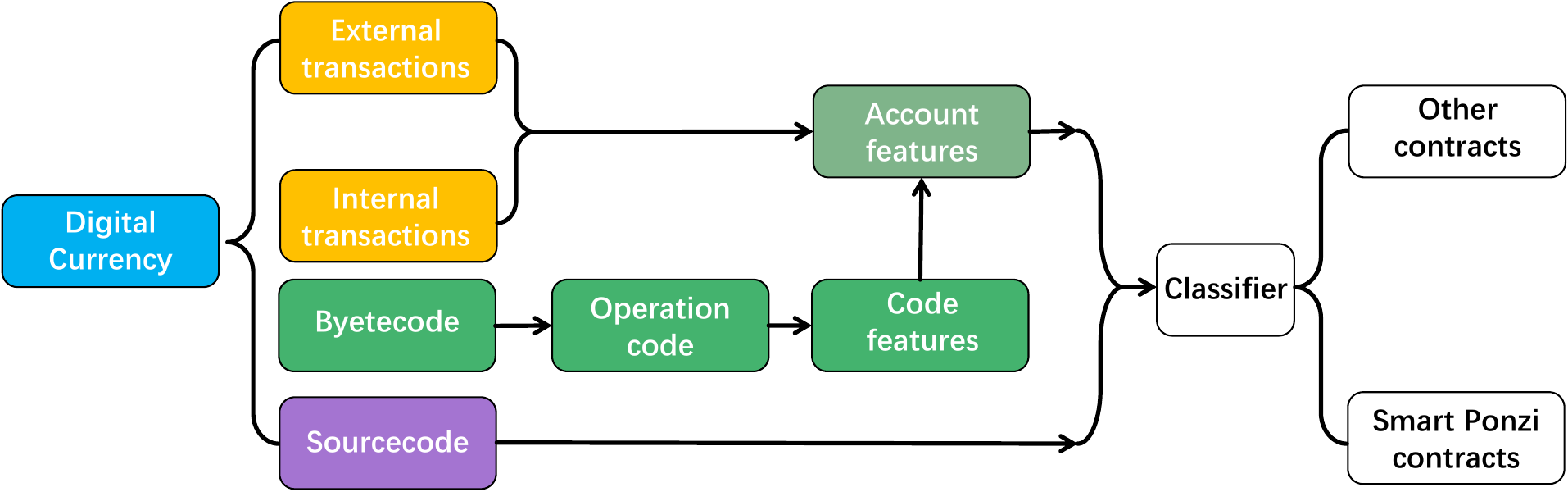}
    \caption{Classification of Smart Ponzi Scheme Detection Methods}
    \label{fig:diagram}
\end{figure}

Currently, smart Ponzi scheme detection methods can be categorized into four main approaches (Figure~\ref{fig:diagram}):  identification based on user behavior analysis, data flow analysis, smart contract bytecode and opcode analysis, and smart contract source code analysis. Each approach offers unique strengths and limitations(Table~\ref{tab:related}), evolving to meet the challenges posed by increasingly sophisticated fraud patterns.

\subsection{Identification Based on User Behavior}
User behavior-based methods examine transaction patterns and interaction anomalies within the blockchain network. For instance, Bartoletti et al.~\cite{bartoletti2020} analyzed 191 active smart Ponzi schemes on Ethereum to uncover their operational mechanisms and impacts. Similarly, Cai et al.~\cite{Cai2018} investigated decentralized applications to establish a foundation for understanding user behavior. While these approaches effectively detect well-known Ponzi patterns, they rely heavily on extensive transaction datasets, making it difficult to recognize emerging or rapidly evolving schemes. More advanced methods, such as the supervised machine learning models introduced by Chen et al.~\cite{Chen2021}, have improved accuracy but depend on large volumes of labeled data. The AI-SPSD model proposed by Fan et al.~\cite{fan2021spsd} introduced unbiased classification strategies, yet practical deployment remains constrained by data imbalance and limited generalization.

\subsection{Identification Based on Data Flow}
Data flow analysis focuses on state transitions and data flow dynamics within smart contracts. Chen et al.~\cite{chen2018detecting} extracted account and code features, employing the XGBoost algorithm to construct a regression tree model for detection. Wang et al.~\cite{Wang2021} built a dataset combining account and contract features, leveraged SMOTE for sample balancing, and trained an LSTM model for classification. Although these methods process large-scale transaction data well, their dependence on high-quality feature engineering and labeled data restricts adaptability. For example, Chen et al.’s TextCNN-transformer approach~\cite{chen2021TextCNN} substantially improved structural understanding through abstract syntax trees and SBT sequences, but it faces challenges regarding computational overhead and real-time deployment.

\subsection{Identification Based on Bytecode and Opcodes Analysis}
Bytecode and opcode-based analysis methods leverage low-level contract features. Zheng et al.~\cite{zheng2023} demonstrated that static bytecode analysis can accurately detect fraudulent behaviors in smart contracts. Although these approaches exhibit high accuracy, traditional techniques such as those by Mason and Escott~\cite{Mason2018} struggle with the increasing complexity and diversity of modern smart contracts. More recent innovations, including the random forest model proposed by Chen et al.~\cite{Chen2020}, have improved detection performance, yet their efficacy still depends on the quality of extracted features. Semantic-oriented methods like SadPonzi~\cite{7} showed promise by analyzing contract bytecode, but handling novel fraud patterns and adapting to the evolving blockchain ecosystem remain significant challenges.

\subsection{Identification Based on Source Code Analysis}
Source code-based approaches directly examine the logic and semantics of smart contracts, capturing fraudulent behaviors more comprehensively than bytecode-level methods. Early work by Mohanta et al.~\cite{Mohanta2018} established a basis for source code-based detection but offered limited real-time monitoring capabilities. Subsequent research, such as Chen et al.~\cite{chen2018detecting}, improved accuracy but encountered difficulties in adapting to code variations. Advanced models like GraphCodeBERT further enhanced feature extraction by leveraging pre-trained code representations and data flow analysis. In addition, Lu et al.~\cite{20} proposed the SourceP method, noting that the required feature data volume grows as the smart Ponzi scheme’s lifespan increases.

\subsection{Discussion}
The aforementioned approaches can also be grouped by whether they rely on dynamic or static data. Dynamic methods emphasize transaction patterns and user behavior, often identifiable only after a contract is deployed on the blockchain. While this post-hoc perspective can capture real user interactions, the fraudulent contract may already be active, posing challenges for timely intervention. In contrast, static analysis leverages contract bytecode or source code before deployment, offering preventive advantages and reducing the risk of feature loss. In particular, source code-based analysis retains richer semantic information than bytecode methods, thereby improving detection precision. This semantic depth facilitates more accurate classification and timely identification, which is critical for mitigating financial losses in smart Ponzi scheme scenarios.

In addition, with the considerable progress of fuzz testing in the field of security\cite{MSurvey2}, it is also worth considering whether to introduce the fuzz testing method for smart Ponzi schemes. And with the emergence of large language models (LLMS), exploring the potential of LLMS in Pontine contract detection has been noticed by some researchers\cite{MSurvey6}; especially projects designed with ChatGPT as the main production tool have accelerated the revolution of artificial intelligence\cite{MSurvey7}. How to better apply these advanced methods to future smart Ponzi scheme detection tasks is also worth our consideration.

\section{Conclusion and Future Work} This paper introduces a new contrastive learning framework, which can effectively learn the semantic information of the data itself, and then identify smart Ponzi scheme contracts. The feasibility of the method is proved by experiments, and it shows that the method has good generalization ability. At the same time, a new method of computing the cosine similarity between multiple vectors is proved and derived. In this method, the similarity between the whole group of vectors is represented by the Angle between an intermediate vector with the same Angle as every vector and any vector.
The future work will optimize our method in terms of the time and the cost.
\section{Acknowledgment} This work was supported by the National Natural Science Foundation of China (U2336204)

\begin{IEEEbiography}[{\includegraphics[width=1in,height=1.25in,clip,keepaspectratio]{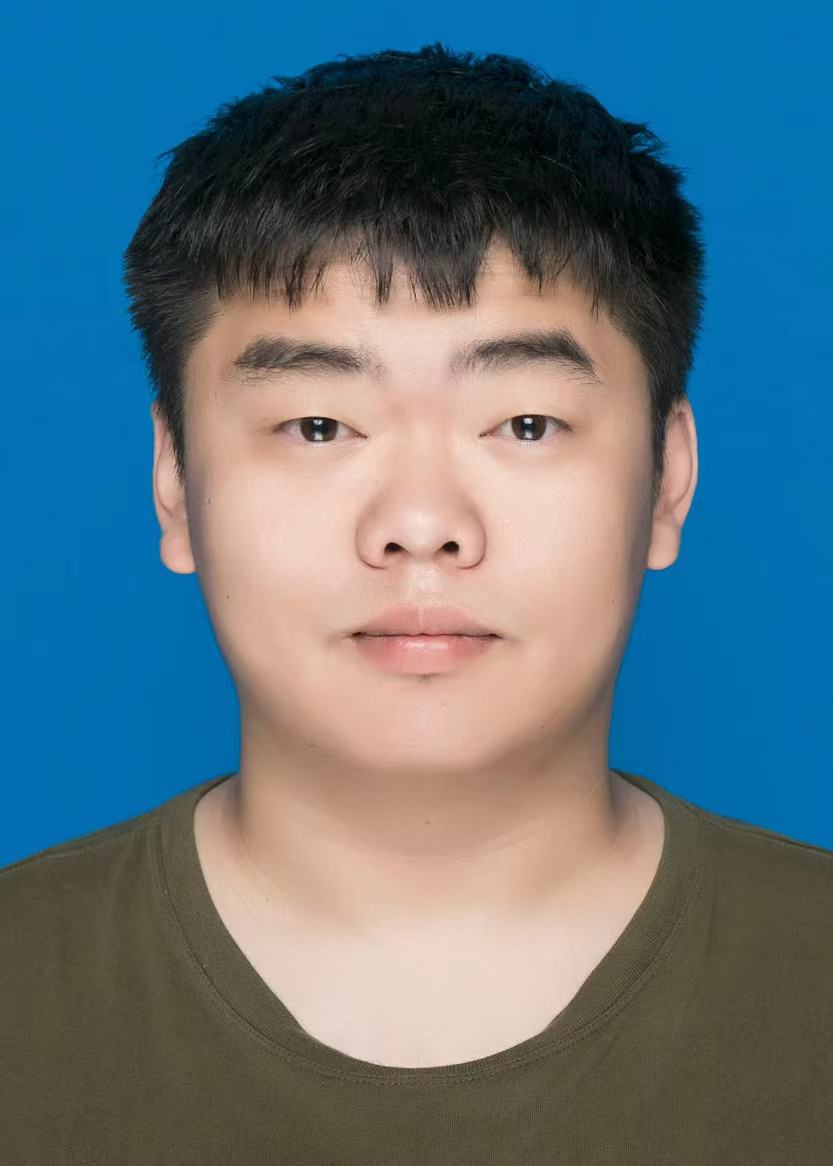}}]{Weijia Yang} graduated from Chengdu University of Information Science and Technology with a master's degree in Mathematics in 2024. He is currently studying for a PhD in engineering at the University of Electronic Science and Technology of China. His research interests include blockchain security, deep learning algorithms, and unsupervised algorithms.
\end{IEEEbiography}
\vspace{-1.1cm}
\begin{IEEEbiography}[{\includegraphics[width=1in,height=1.25in,clip,keepaspectratio]{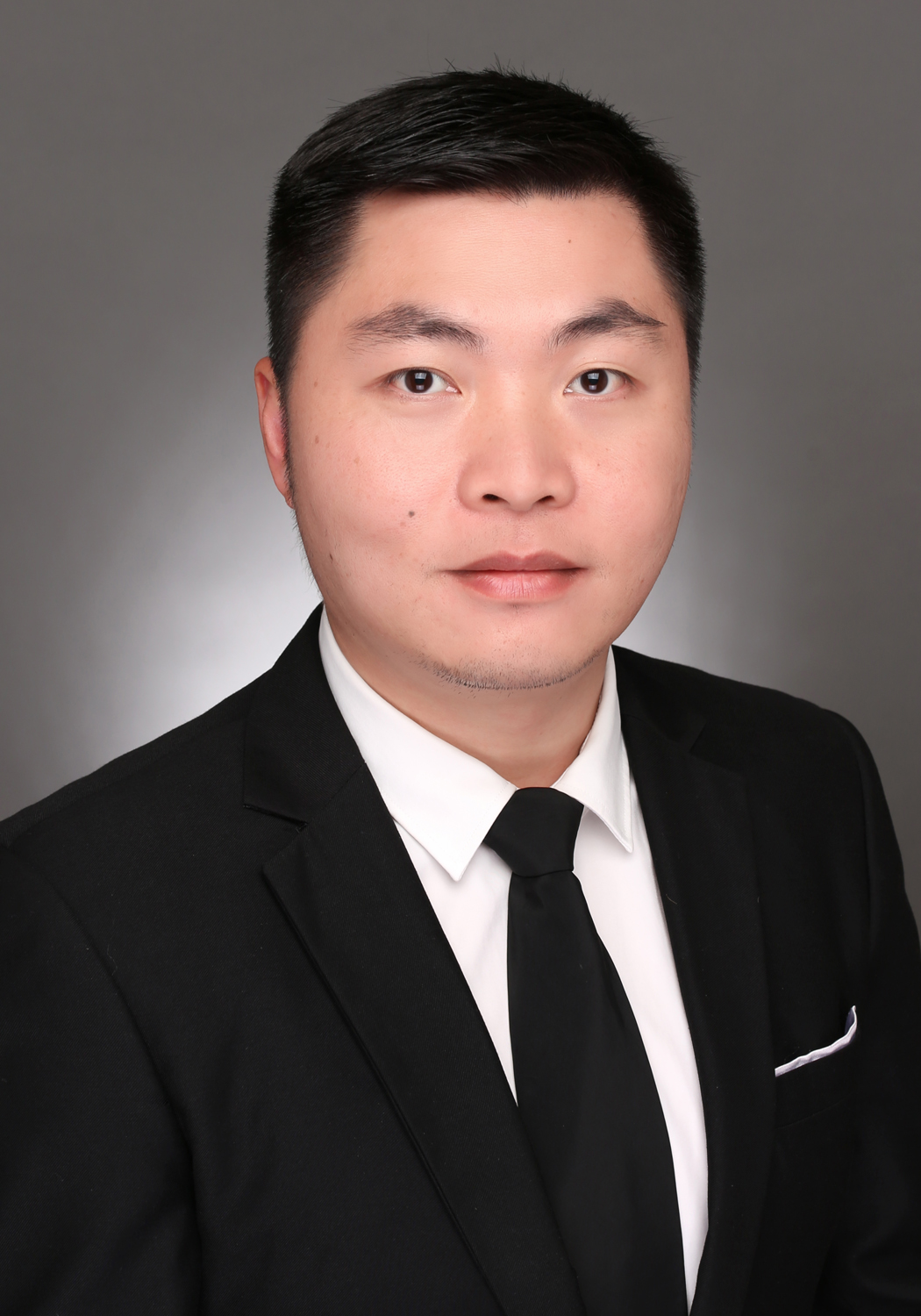}}]{Tian Lan} graduated from the University of Electronic Science and Technology of China with a doctorate in engineering in 2009, and currently works as a researcher in the School of Cyberspace Security of the University of Electronic Science and Technology of China. His research interests include blockchain security, natural language processing, semantic enhancement, etc.
\end{IEEEbiography}
\vspace{-1.1cm}
\begin{IEEEbiography}[{\includegraphics[width=1in,height=1.25in,clip,keepaspectratio]{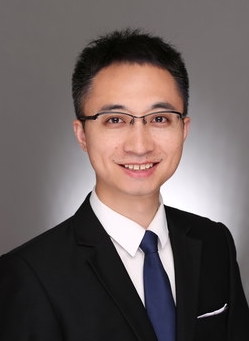}}]{Leyuan Liu} received the B.E. degree in Information Security from University of Electronic Science and Technology of China, in 2006, and the Ph.D. in Advanced Information Technology from Kyushu University, Japan, in 2014. He is currently a Research Associate at the School of Information and Software Engineering, University of Electronic Science and Technology of China. His research interests include graph learning, information dissemination, social network analysis, and blockchain security.
\end{IEEEbiography}

\vspace{-1.1cm}

\begin{IEEEbiography}[{\includegraphics[width=1in,height=1.25in,clip,keepaspectratio]{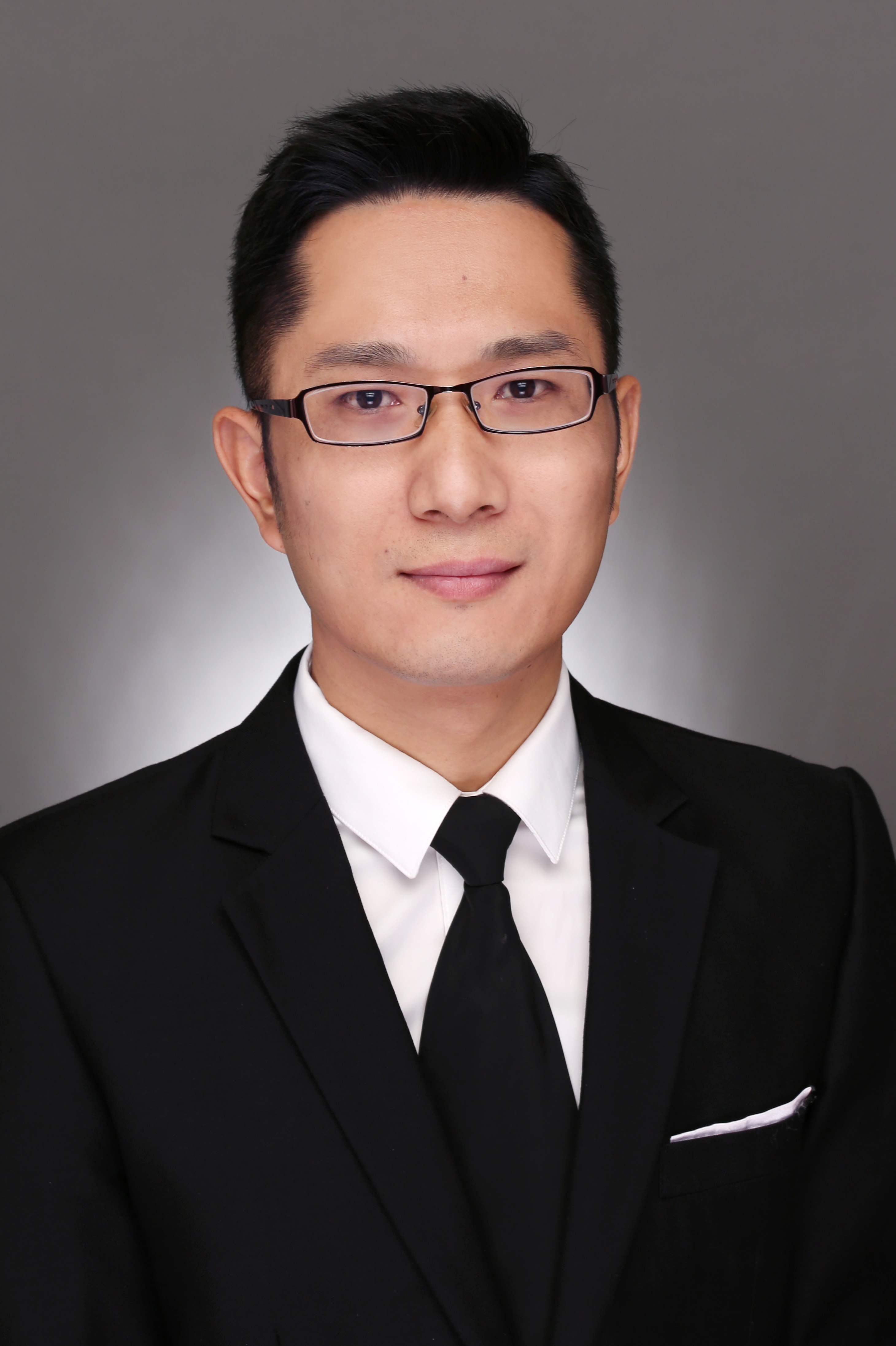}}]{Wei Chen} obtained his Bachelor's degree in Computer Science and Engineering from University of Electronic Science and Technology of China from September 1997 to July 2001. He later pursued a Master's degree in Computer Application Technology at the same university from March 2001 to July 2004. From March 2004 to December 2010, he studied for a Ph.D. in Information and Communication Engineering.
\end{IEEEbiography}

\vspace{-1.1cm}

\begin{IEEEbiography}[{\includegraphics[width=1in,height=1.25in,clip,keepaspectratio]{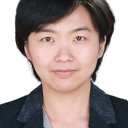}}]{Tianqing Zhu} is a professor at City University of Macau, before that, she was a lecturer at Deakin University in Australia, an associate professor at the University of Technology Sydney, and a professor at China University of Geosciences (Wuhan). She was also a College of Expert (CoE) in Australian Research Council. She has led and participated in Eight Australian Research Council projects with a total research funding of more than 4 million Australian dollars. She has published 300 SCI papers in total. She serves as PC Member of the International Conference on Security CCS 2025, PC Member of the International Conference on Artificial Intelligence AAAI, IJCAI, and Associate Editor of 3 SCI journals. She has committed to the field of artificial intelligence security, focusing on key scientific issues such as intelligent model security attack and defense, data privacy, and the relationship between security and fairness, and improving the security, privacy protection and output fairness of intelligent models.
\end{IEEEbiography}

\vspace{-1.1cm}

\begin{IEEEbiography}[{\includegraphics[width=1in,height=1.25in,clip,keepaspectratio]{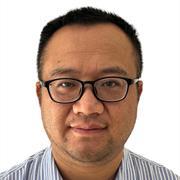}}]{Sheng Wen} received the Ph.D. degree in Computer Science from the School of Information Technology, Deakin University, Australia, in 2015. He is currently a Senior Lecturer at Swinburne University of Technology. His focus is on modeling of virus spread, information dissemination, and defense strategies for the Internet threats. He is also interested in the techniques information sources in networks.
\end{IEEEbiography}
\vspace{-1cm}
\begin{IEEEbiography}[{\includegraphics[width=1in,height=1.25in,clip,keepaspectratio]{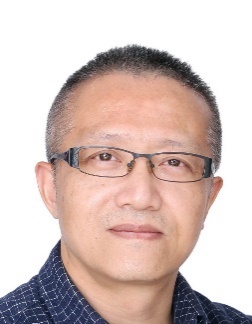}}]{Xiaosong Zhang} received the B.S. degree in dynamics engineering from
Shanghai Jiaotong University, Shanghai, in 1990, and the M.S. and Ph.D.
degrees in computer science from the University of Electronic and Technology of China (UESTC), Chengdu, in 2011. He has worked on numerous projects in both research and development roles. He is currently an Associate Director with the National Engineering Laboratory of Big Data application to improving the Government governance capacity in China.
\end{IEEEbiography}

\end{document}